\documentclass[12pt]{article}
\usepackage{amsmath,amssymb}
\usepackage{physics}
\usepackage{geometry}
\newtheorem{theorem}{Theorem}

\title{\textbf{\mbox{On the Cocycle Structure of}\vspace{0.4cm}\\ 
\mbox{the Boltzmann Distribution}}}
\vspace{1.5cm}
\author{
Chuan-Tsung Chan, Chan-Yi Chang, Zhong-Tang Wu
\vspace{0.4cm}\\
\small Department of Applied Physics, Tunghai University,
 Taichung 40704, Taiwan
}

\date{}

\begin{document}

\maketitle









\begin{abstract}
Based on a cocycle structure, we identify a new derivation of the Boltzmann distribution for finite energy-level systems from the maximal entropy principle (MEP). Our approach does not rely on the method of the Lagrange multiplier, and it provides a more transparent way to understand the dependence on the energy levels of the temperature \( T = 1/\beta \) for the equilibrium distribution. Finally, we make two curious observations associated with our derivations.
\end{abstract}




\section{Introduction}

The main theme of statistical mechanics is to embody the microscopic dynamics in the context of probability theory. In particular, the macroscopic equilibrium state corresponds to the set of microscopic states with maximal degeneracy. The large number limit, i.e., the Stirling formula of the gamma function, of the degeneracy naturally gives rise to the definition of the statistical entropy. The definition and principle of the maximal entropy lead to various probability distribution functions subject to given constraints. In the simplest case, for the canonical ensemble, we fix the expectation value of the energy and deduce the Boltzmann distribution. 
\\

In the standard textbook description \textbf{[1]}, \textbf{[2]}, to solve the optimization problem with constraints, one typically resorts to the method of Lagrange multiplier, in which this avoids the difficulty of obtaining explicit solutions of the constraints and in general preserves the symmetry of the original problem. However, from such an approach it may not be easy to identify the Lagrange multipliers as functions of parameters and the type of critical point (maxima or minima) is not apparent in the Lagrange multiplier method. For these reasons, we opt for a different approach of the maximal entropy problem by solving the constraints directly. Our approach is based on a reformulation of the setup in a geometrical way, and it overcomes the weakness of the Lagrange multiplier method.\\

This paper is organized as follows. For the pedagogical reason we divide the discussion/derivations in two parts: In the first part (Sec.2 and Sec.3), we use the simplest example - three-level system to illustrate the key idea of our approach. Then we generalize to the case of arbitrary $N$-level systems in Sec.4. Incidentally, our derivations lead to some interesting observations which are summarized in Sec.4.\\

\section{An illustrative example: three-level system}


To begin with, we treat the occupation probabilities: \( p_k \), with \( k = 1, 2, 3 \), 
subject to the following constraints:
\begin{equation}
p_k > 0 \quad \forall k, \qquad
\sum_{k=1}^{3} p_k = 1, \qquad
\sum_{k=1}^{3} p_k E_k = E .
\end{equation}
In the last expression, the energy expectation value, a sum of level energies, $E_k$, weighted by the probability distribution \( p_k \), is fixed to be \( E \). Our strategy for solving the optimization problem associated with the maximal entropy principle (MEP) is to treat the variables as coordinates of vectors in an $N$-dimensional vector space equipped with a Euclidean metric. In this formulation, the probability normalization and the energy expectation value constraints become linear conditions on the probability vector.

We define the vectors
\begin{equation}
\vec{I} := (1,1,1), \qquad
\vec{e} := (e_1, e_2, e_3), \hspace{0.5cm}e_k := E_k/E,  \qquad
\vec{p} := (p_1, p_2, p_3).
\end{equation}
The constraints, Eq.(1), can then be written compactly as
\begin{equation}
\vec{I} \cdot \vec{p} = 1, \qquad
\vec{e} \cdot \vec{p} = 1 .
\end{equation}
We further identify another vector
\begin{equation}
\vec{f} := \vec I \times \vec e = (e_{32}, e_{13}, e_{21}),
\end{equation}
where the reduced-energy differences are defined as
\begin{equation}
e_{jk} := e_j - e_k, \hspace{0.5cm} j,k = 1,2,3 .
\end{equation}
One can show that
\begin{equation}
\vec I\cdot \vec f = e_{32}+e_{21}+e_{13}=0  \hspace{0.5cm}\text{and} \hspace{0.5cm}\vec e \cdot \vec f=e_{32}e_1+e_{13}e_2+e_{21}e_3=0 .
\end{equation}
Moreover, the determinant
\begin{equation}
\det(\vec I,\vec e,\vec f)
=
\begin{vmatrix}
1 & 1 & 1\\
e_1 & e_2 & e_3\\
e_{32} & e_{13} & e_{21}
\end{vmatrix}
=
(e_1-e_2)^2+(e_2-e_3)^2+(e_3-e_1)^2
\neq 0
\end{equation}
for non-degenerate energy levels. Therefore, the set $\{\vec I,\vec e,\vec f\}$ forms a linearly independent complete basis for the vector space $\mathbb R^3$. Consequently, the probability vector, $\vec p=(p_1,p_2,p_3)$, can be expanded as
\begin{equation}
\vec p=  a \vec I+ b \vec e+ x \vec f .
\end{equation}
The coefficients $a$ and $b$ can be determined by taking projections with respect to the basis vectors. Explicitly, one finds
\begin{equation}
\vec I\cdot\vec p = 3  a +3 \overline{e}  b = 1 , \qquad
\vec e\cdot\vec p = 3  \overline{e}  a +3\overline{e^2}  b  = 1 .
\end{equation}
Here the average energy, $ \overline{e}:= (e_1 + e_2 + e_3) /3 , \hspace{0.2cm}\text{and the average energy-squared,}\\\hspace{0.2cm}\overline{e^2}:= (e_1^2 + e_2^2 + e_3^2) /3 $.
Solving these equations yields the unique solution
\begin{equation}
\begin{pmatrix}
a \\[4pt]
b
\end{pmatrix}
=
\frac{1}{3}
\begin{pmatrix}
1 & \overline{e} \\
\overline{e} & \overline{e^2}
\end{pmatrix}^{-1}
\begin{pmatrix}
1 \\[4pt]
1 \label{coeff1}
\end{pmatrix}
=
\frac{1}{3\!\left(\overline{e^2}-\overline{e}^{\,2}\right)}
\begin{pmatrix}
\overline{e^2}-\overline{e} \\[4pt]
1-\overline{e}
\end{pmatrix}.
\end{equation}
In this way, the two constraints are explicitly solved, and the probability vector $\vec p$ is parametrized by a single variable $x$.\\

The entropy function associated with the three-level system is defined as
\begin{equation}
S(\vec p):=-\sum_{k=1}^3 p_k\ln p_k > 0.
\end{equation}
Upon substituting the parametrized probability vector,
\begin{equation}
\vec p(x)=  a \vec I+ b \vec e+ x \vec f  ,  \hspace{0.5cm}\frac{d}{dx}\vec p(x) = \vec f,
\end{equation}
the maximal entropy problem is reduced to a single-variable optimization problem. We define
\begin{equation}
s(x):=S(\vec p(x)) =-\sum_{k=1}^3 p_k(x)\ln p_k(x).
\end{equation}

\begin{equation}
0=\frac{d s}{d x}
=-\sum_{k=1}^N\left[
\frac{d p_k}{d x}\ln p_k
+\frac{d p_k}{d x}
\right] =: \vec f \cdot \vec q(x). \label{critqeq1}
\end{equation}
Here we define the log-probability vector, 
\begin{equation}
\vec q(x):= \left(-\ln p_1{(x)}, -\ln p_2{(x)}, -\ln p_3{(x)}\right),
\end{equation}
which naturally emerges in the entropy variation. Note that
\begin{equation}
    \frac{d}{dx}\vec q(x) = -\left(\frac{e_{32}}{p_{1}(x)}, \frac{e_{13}}{p_{2}(x)}, \frac{e_{21}}{p_{3}(x)}\right). \label{diffq1}
\end{equation}

Since we have shown that the set $\{\vec I,\vec e,\vec f\}$ forms a complete basis for $\mathbb R^3$, the critical point condition, Eq.(\ref{critqeq1}), implies that the log-probability vector $\vec q$ must lie in the subspace spanned by $\vec I$ and $\vec e$. That is,
\begin{equation}
\vec q_* := \vec q(x_*) = \alpha\,\vec I + \beta\,\vec e .
\label{critq1}
\end{equation}
Equivalently, the critical point solution of the maximal entropy problem is uniquely given by the Boltzmann distribution,
\begin{equation}
\vec p_* := \vec p(x_*) = (e^{-\alpha - \beta e_1},\ e^{-\alpha - \beta e_2}, \ e^{-\alpha - \beta e_3}), \hspace{0.5cm}. \label{critqp1}
\end{equation}
Furthermore, one can show that the critical solution for the three-level system indeed corresponds to a maximal value of the entropy function, using Eq.(\ref{diffq1}),

\begin{equation}
\frac{d^2 s}{dx^2}
=\vec f \cdot \frac{d\vec q(x)}{dx} \Bigg|_{\vec q = \vec q_*}
= 
-\frac{e_{32}^2}{p_1}
-
\frac{e_{13}^2}{p_2}
-
\frac{e_{21}^2}{p_3}
< 0 .
\end{equation}

\section{Solving the parameters in terms of the energy levels}

From the solution obtained, Eq.(\ref{critq1}), we identify the partition function,
\begin{equation}
Z := e^\alpha = \sum_{k=1}^{3} e^{-\beta e_k} =: \sum_{k=1}^{3} \xi ^{e_k}, \hspace{0.5cm} \xi := e^{-\beta}.
\end{equation}

On the other hand, we can solve $x_*$ and $\xi$ from the algebraic equations, Eq.(\ref{critqp1}),
\begin{eqnarray}
& &p_1
=
a+be_1+x_*e_{32}
=
\frac{1}{Z} e^{-\beta e_1} =: \frac{1}{Z}\xi^{e_1}.\nonumber\\
& &p_2
=
a+be_2+x_*e_{13}
=
\frac{1}{Z} e^{-\beta e_2} =: \frac{1}{Z}\xi^{e_2}. \nonumber\\
& &p_3
=
a+be_3+x_*e_{21}
=
\frac{1}{Z} e^{-\beta e_3} =: \frac{1}{Z}\xi^{e_3}. \label{pcome}
\end{eqnarray}
Here $a$,$b$ are given in Eq.(\ref{coeff1}), and only two out the three equations above are independent.\\

To solve for $x_*$, we notice that the occupation probabilities satisfy a cocycle condition,
\begin{equation}
    e_{32}q_1 + e_{13}q_2 + e_{21}q_3 = 0 \hspace{0.3cm}
    \Leftrightarrow \hspace{0.3cm} p_1^{e_{32}}\cdot p_2^{e_{13}}\cdot p_3^{e_{21}} = 1. 
\end{equation}
Consequently, the critical value $x_*$ satisfies the following equation,
\begin{equation}
    (a+be_1+x_*e_{32})^{e_{32}}(a+be_2+x_*e_{13})^{e_{13}}(a+be_3+x_*e_{21})^{e_{21}} = 1.
\end{equation}
Substituting the solution of $x_*$ into any component of the probabilities vector, Eq.(\ref{pcome}), we obtain the solution of $\xi := e^{-\beta}$.

Thus, as promised, we have derived exact expressions showing
$\alpha$ (corresponding to the partition function) and $\beta$
(corresponding to the inverse temperature) as implicit functions
of the reduced-energy levels, $\{e_1, e_2, e_3\}$.

\section{The case of $N$ level systems}
We hope that, through our simple illustration of the three-level system,
one can be convinced that the generalization to the $N$-level system
is straightforward. In the following, we sketch the necessary steps of such a generalization.

To begin with, we shall treat all variables as vectors in $\mathbb{R}^N$.
For instance,

\begin{itemize}
  \item constant vector:
  \begin{equation}
  \vec I := (1, 1, \ldots, 1)
  \end{equation}

  \item reduced energy-level vector:
  \begin{equation}
  \vec e := (e_1, e_2, \ldots, e_N) , \hspace{0.2cm} e_k := E_k / E
  \end{equation}

  \item probability vector:
  \begin{equation}
  \vec p := (p_1, p_2, \ldots, p_N)
    \end{equation}
    
  \item log-probability vector: 
  \begin{equation}
      \vec q := (-\ln p_1,\,-\ln p_2,\, \ldots, -\ln p_N)
  \end{equation}

  
\end{itemize}
In this geometrical formulation, the probability normalization and energy expectation value constraints become hyperplanes in
$\mathbb{R}^N$,
\begin{equation}
\vec I \cdot \vec p = 1, \qquad
\vec e \cdot \vec p = 1 .
\end{equation}

To construct a complete set of basis vectors for $\mathbb{R}^N$,
we supplement $\vec I$ and $\vec e$ with the following $N-2$ vectors:
\begin{eqnarray}
    & &\vec f_1 := (e_{32}, e_{13}, e_{21}, \vec 0_{N-3}), \nonumber\\
    & &\vec f_2 := (0, e_{43}, e_{24}, e_{32}, \vec 0_{N-4}), \nonumber\\
    & &\vec f_k :=
(\vec 0_{k-1}, e_{k+2\,k+1}, e_{k\,k+2}, e_{k+1\,k}, \vec 0_{N-k-2}) .
\end{eqnarray}
It is clear that
\begin{equation}
\vec I \cdot \vec f_k = \vec e \cdot \vec f_k = 0 ,
\qquad k = 1,2,\ldots ,N-2,
\end{equation}
and $\{\vec f_1, \vec f_2, \ldots, \vec f_{N-2}\}$ forms a set of linearly independent vectors. Consequently, the full space $\mathbb{R}^N$ can be decomposed as a direct sum of
two orthogonal complements,
\begin{equation}
\mathbb{R}^N
=
\mathrm{span}\{\vec I, \vec e\}
\oplus
\mathrm{span}\{\vec f_1, \vec f_2, \ldots, \vec f_{N-2}\}.
\end{equation}

We now expand the probability vector $\vec p$ in terms of this set of basis vectors,
\begin{equation}
\vec p
=
a\,\vec I
+
b\,\vec e
+
\sum_{k=1}^{N-2} x_k \vec f_k . 
\end{equation}
By taking projection with respect to $\vec I$ and $\vec e$, We have
\begin{equation}
1
=
\vec I \cdot \vec p
=
N a + N \overline{e} b ,
\qquad
\overline{e} := \frac{1}{N}\sum_{k=1}^{N} e_k , \label{cons21}
\end{equation}

\begin{equation}
1
=
\vec e \cdot \vec p
=
N \overline{e} a + N \overline{e^2}\ b ,
\qquad
\overline{e^2} := \frac{1}{N}\sum_{k=1}^{N} e_k^2 .\label{cons22}
\end{equation}
The solutions of $a,b$ are,
\begin{equation}
\begin{pmatrix}
a \\[4pt]
b
\end{pmatrix}
=
\frac{1}{N}
\begin{pmatrix}
1 & \overline{e} \\
\overline{e} & \overline{e^2}
\end{pmatrix}^{-1}
\begin{pmatrix}
1 \\[4pt]
1 \label{coeff2}
\end{pmatrix}
=
\frac{1}{N\!\left(\overline{e^2}-\overline{e}^{\,2}\right)}
\begin{pmatrix}
\overline{e^2}-\overline{e}\, \\[4pt]
1-\overline{e}
\end{pmatrix}.
\end{equation}

Having solved the constraints, Eqs.~(\ref{cons21}),(\ref{cons22}), we obtain a ``covariant''
parametrization of the probability vector in terms of $N-2$ variables,
\begin{equation}
\vec p = \vec p(x_1, x_2, \ldots, x_{N-2}). \label{prob2}
\end{equation}

The entropy function for a $N$-level system is defined as
\begin{equation}
S(\vec p)
:=
- \sum_{k=1}^{N} p_k \ln p_k .
\end{equation}
By substituting the probability vector, Eq.~(\ref{prob2}), into the entropy function, we define
\begin{eqnarray}
s(x_1, x_2, \ldots, x_{N-2})&:=&S(\vec p(x_1, x_2, \ldots, x_{N-2}))\nonumber\\& &\nonumber\\
&=&-\sum_{k=1}^N p_k(x_1, x_2, \ldots, x_{N-2})\ln p_k(x_1, x_2, \ldots, x_{N-2}).
\end{eqnarray}

The extremum condition is given by
\begin{equation}
0=\frac{\partial s}{\partial x_j}
=-\sum_{k=1}^N\left[
\frac{\partial p_k}{\partial x_j}\ln p_k
+\frac{\partial p_k}{\partial x_j}
\right] =: \vec f_j \cdot \vec q  \label{critqeq2}
\end{equation}

Here we introduce the log-probability vectors,
\begin{equation}
q_k:=-\ln p_k \geq 0,  \hspace{0.5cm} \vec q := (-\ln p_1,\,-\ln p_2,\,\ldots,-\ln p_N),
\end{equation}
and we have a cocycle condition, 
\begin{equation}
e_{k+2 \hspace{0.05cm} k+1} \ln p_k + e_{k \hspace{0.05cm} k+2} \ln p_{k+1} + e_{k+1 \hspace{0.05cm} k} \ln p_{k+2} = 0. 
\end{equation}

Since we have shown that the set $\{\vec I,\vec e,\vec f_1, \vec f_2, \ldots,\vec f_{N-2}\}$ forms a complete basis for $\mathbb R^N$, the critical point condition, Eq.(\ref{critqeq2}), implies that the log-probability vector must lie in the subspace spanned by $\vec I$ and $\vec e$. That is,
\begin{equation}
\vec q_* = \alpha\,\vec I + \beta\,\vec e . \label{critq21}
\end{equation}

Equivalently, the critical point solution of the maximal entropy problem is uniquely given by the Boltzmann distribution,
\begin{equation}
p_k = e^{-\alpha - \beta e_k}, \qquad k = 1, 2, \ldots, N. \label{critq22}
\end{equation}

From this solution, we see that, for arbitrary three levels, $e_j$, $e_k$, $e_l$, the Boltzmann distribution satisfies the general cocycle condition,

\begin{theorem}
\begin{equation}
p_j^{e_{l \hspace{0.05cm} k}} p_l^{e_{k \hspace{0.05cm} j}} p_k^{e_{j \hspace{0.05cm} l}} = 1.
\end{equation}
\end{theorem}

Furthermore, we observe that this cocycle condition has an interesting implication:
\begin{theorem}
If the energy levels of a finite system are all rational numbers, then the Boltzmann probabilities are all algebraic numbers.
\end{theorem}
 
To show that the critical point given by Eqs.(\ref{critq21}),(\ref{critq22}), indeed corresponds to the maximal entropy, 
we compute the Hessian matrix of the entropy function.
The second derivative of the entropy with respect to the coordinates $x_j$ and $x_k$ reads
\begin{eqnarray}
\frac{\partial^2 S}{\partial x_k \partial x_j}
&=&\frac{\partial}{\partial x_k} \left( \vec f_j \cdot \vec{q} \right)
= \vec {f}_j \cdot \frac{\partial \vec{q}}{\partial x_k}
= - \left(
\frac{f_{j1} f_{k1}}{p_1}
+ \frac{f_{j2} f_{k2}}{p_2}
+ \cdots
+ \frac{f_{jN} f_{kN}}{p_N}
\right) \nonumber\\
&=& - \vec{h}_j \cdot \vec{h}_k
=: - g_{jk}, \qquad j,k = 1, 2, \ldots, N-2 \label{hessian}
\end{eqnarray}
Here we have used 
\begin{equation}
    \frac{\partial\vec q}{\partial x_j} = \left(-\frac{f_{j1}}{p_1}, -\frac{f_{j2}}{p_2}, \ldots, -\frac{f_{jN}}{p_N}\right),\hspace{0.5cm} j = 1,2,\ldots,N-2.
\end{equation}
and define a new set of vectors,
\begin{equation}
      \vec h_j := \left(-\frac{f_{j1}}{\sqrt p_1}, -\frac{f_{j2}}{\sqrt p_2}, \ldots, -\frac{f_{jN}}{\sqrt p_N}\right) \label{skewprob},
\end{equation}
together with the metric tensor, $g_{jk}$. We thus show that the Hessian matrix of the entropy function, Eq.(\ref{hessian}), is equal to the negative metric tensor matrix of $\{ \vec{h}_j \}_{j=1}^{N-2}$.
We note that all the principal minor determinants of the negative Hessian, $-\frac{\partial^2 S}{\partial x_k \partial x_j}$, are determinants of associated with smaller subspaces expanded by $\{ \vec h_j \}_{j=1}^{m}$, $ 1 \leq m \leq N-2$, metric tensor matrices. Hence, we see that they are all positive, and this proves that, via the Sylvester's criterion \textbf{[3]}, the critical point is indeed the maximal of the entropy function.

\section{Summary and Conclusion}
In this paper, we have devised an alternative approach to derive the Boltzmann distribution based on the maximal entropy principle (MEP). 
Based on a geometrical formulation, we solve the probability normalization and energy expectation constraints via the standard procedure in linear algebra. 
Our approach incorporates the physical variables, namely the partition function ($\alpha = \ln Z$) and the inverse temperature $(\beta = 1/T)$, as parts of the unknown variables (together with $N-2$ complementary ``coordinates''). 
This results in a change of variables for the equilibrium probability distributions, from $(p_1, p_2, \ldots, p_N)$ to $(\alpha, \beta, x_1, \ldots, x_{N-2})$. Our derivation not only demonstrates the energy-level dependence on the inverse temperature, but also provides explicit proof that the critical probability distributions are indeed maxima of the entropy function subject to the relevant constraints.

As a final remark, it may appear that we do not really need to specify the explicit form of the complementary vectors
$\{ \vec f_j \}_{j=1}^{N-2}$ in order to show that the logarithmic Boltzmann distribution, $\vec q(x_*)$, is a linear combination of
$\alpha$ and $\beta$. 
However, our choice does help in solving $\beta$ and $\alpha$ in terms of the reduced energy-level difference, $e_{jk} := e_j - e_k$ ,
in the form of a cocycle structure, \textbf{Theorem 1}. Furthermore, the consequence implied by this cocycle structure, \textbf{Theorem 2}, may be of interest to the study of number theory.

It is possible to extend our approach to the case of the grand-canonical ensemble by introducing a new basis vector corresponding to the particle number / chemical potential distribution. 
At this moment, it is not completely clear whether there exists a similar argument that can generate the Bose--Einstein and/or the Fermi--Dirac distributions in the quantum statistical context.

Our work is supported by the grants of the National Science and Technology Council (NSTC) of Taiwan, under the following budgets: 114-2112-M-029-003, 114-2112-M-029-001, and 113-2112-M-029-002.


\end{document}